\documentclass[journal]{IEEEtran}
\usepackage{latexsym}
\usepackage{graphicx}
\usepackage{float}
\usepackage{graphicx,psfrag,amsmath,amsfonts,verbatim}
\usepackage{amssymb}
\usepackage[small,bf]{caption}
\usepackage{indentfirst}
\usepackage{bm}
\usepackage[ruled,vlined,linesnumbered]{algorithm2e}
\usepackage{cite}
\usepackage[T1]{fontenc}
\usepackage[utf8]{inputenc}
\usepackage{authblk}
\usepackage{setspace}
\usepackage{fancyhdr}
\usepackage{multirow}
\usepackage{textcomp}
\def\BibTeX{{\rm B\kern-.05em{\sc i\kern-.025em b}\kern-.08em
    T\kern-.1667em\lower.7ex\hbox{E}\kern-.125emX}}
\begin{document}
\bibliographystyle{IEEEtran}
\title{Intelligent Reflecting Surface Enhanced D2D Cooperative Computing}
\author{Sun~Mao,~
        Xiaoli Chu,
        Qingqing Wu,
        Lei Liu,
        and Jie Feng
\IEEEcompsocitemizethanks{This work was supported in part by the National Natural Science
Foundation of China (No. 62001357, 62001028, 61903384), in part by the Guangdong Basic and Applied Basic Research Foundation (No.2020A1515110496, 2020A1515110079), in part by the Fundamental Research Funds for the Central Universities (XJS210107, XJS210105, XJS211513), in part by FDCT 0108/2020/A, and the Open Research Fund of National Mobile Communications Research Laboratory, Southeast University (No. 2021D15). \emph{(Corresponding author: Qingqing Wu and Lei Liu)}.\\
\mbox{~~~}Sun Mao is with the College of Computer Science, Sichuan Normal University, Chengdu, 610101, China. (email: sunmao@sicnu.edu.cn).\\
\mbox{~~~}Xiaoli Chu is with the Department of Electronic
and Electrical Engineering, The University of Sheffield, U.K. (email: x.chu@sheffield.ac.uk).\\
\mbox{~~~}Qingqing Wu is with the State Key Laboratory of Internet of Things for Smart City, University of Macau, Macau, 999078, and also with the National Mobile Communications Research Laboratory, Southeast University, Nanjing 210096, China (email: qingqingwu@um.edu.mo).\\
\mbox{~~~}Lei Liu and Jie Feng are with the State Key Laboratory of Integrated Service Networks, Xidian University,
Xi’an 710071, China (e-mail: tianjiaoliulei@163.com; jiefengcl@163.com).
 }}

\maketitle

\begin{abstract}
This paper investigates a device-to-device (D2D) cooperative computing system, where an user can offload part of its computation task to nearby idle users with the aid of an intelligent reflecting surface (IRS). We propose to minimize the total computing delay via jointly optimizing the computation task assignment, transmit power, bandwidth allocation, and phase beamforming of the IRS. To solve the formulated problem,  we devise an alternating optimization algorithm with guaranteed convergence. In particular, the task assignment strategy is derived in closed-form expression, while the phase beamforming is optimized by exploiting the semi-definite relaxation (SDR) method. Numerical results demonstrate that the IRS enhanced D2D cooperative computing scheme can achieve a much lower computing delay as compared to the conventional D2D cooperative computing strategy.
\end{abstract}
\IEEEpeerreviewmaketitle
\section{Introduction}
Recently, intelligent reflecting surface (IRS) has been envisioned as an innovative technique for beyond fifth-generation (B5G) communication systems, due to its ability to reconfigure the wireless propagation environment \cite{8811733}. In particular, IRS is composed of a large number of passive reflecting elements, which can intelligently reflect incident signals to enhance signal power or suppress co-channel interference by adjusting their amplitudes and phases \cite{8970580}, \cite{9133435}. Different from traditional relay communications, IRS not only can operate in full duplex mode naturally without suffering from undesired self-interference, but also can greatly reduce power consumption and hardware cost by using passive reflecting elements \cite{8910627}.

Meanwhile, mobile edge computing (MEC) has been leveraged to enable a variety of computation-intensive and delay-sensitive applications, such as virtual reality (VR), autonomous driving, interactive online games, and so on \cite{8937028}, \cite{8318578}, \cite{8941121}, \cite{8168252}.  However, the quality of computation offloading service cannot be ensured when the communication link between the user and the AP is blocked occasionally by moving objects in the surrounding environment or due to user mobility \cite{9279326}, \cite{liu2019vehicular}, \cite{liu2020intelligent}. To address this issue, the authors in \cite{9133107} and \cite{9270605} introduced IRS into MEC systems and investigated the latency minimization and computational bits maximization problems, respectively.

In addition to equipping the edge server at access points (APs), device-to-device (D2D) cooperative computing is another promising solution to support computation-intensive services by utilizing the spare computation resources at end users. In particular, IRS has great potential to enhance D2D cooperative computing by establishing high-rate composite links among end users, especially in the scenario that direct communication links between end users are blocked. Different from the edge server with powerful computing capability, user devices each have limited computing capability and energy, it is thus essential to design the task offloading and resource management strategy in D2D cooperative computing, considering both the user computing capability and channel state between cooperative users. However, the methodology for AP-based MEC in \cite{9133107}, \cite{9270605} cannot be applied in the IRS-assisted D2D cooperative
computing. Furthermore, prior works in \cite{8779699}, \cite{9224971} studied the energy-efficient D2D cooperative computing strategy, but they ignored the benefits of IRS to task offloading.

Motivated by these observations, this paper investigates the task offloading and resource management for an IRS-assisted D2D cooperative computing system, which consists of a source node and multiple helper nodes. The source node can offload part of its computation task to nearby helper nodes with the assistance of an IRS. In this scenario, we aim at minimizing the computing delay of source node by optimally designing the computation task assignment, transmit power, bandwidth allocation, and phase beamforming of the IRS. In order to solved the formulated non-convex joint optimization problem, we first derive the optimal computation task assignment in closed-form expression and then obtain the optimal transmit power and bandwidth allocation, as well as the IRS phase beamforming by exploiting the convex optimization theory and SDR method. Simulation results illustrate the delay reduction achieved by our proposed IRS-assisted D2D cooperative computing, when compared to other benchmark methods.

\emph{Notations: }For a vector $\mathbf{x}$,  its transpose, Hermitian transpose and diagonalization are denoted by $\mathbf{x}^T$, $\mathbf{x}^H$ and  diag($\mathbf{x}$), respectively. For a matrix $\mathbf{M}$, its trace is denoted by Tr($\mathbf{M}$), while $[\mathbf{M}]_{mn}$ indicates its element in the $m$-th row and $n$-th column.
\section{System Model}
As shown in Fig. 1, we consider an IRS-assisted D2D cooperative computing system, which includes a single-antenna source node and $K$ single-antenna helper nodes denoted by $\mathcal{K}=\{1,2,\cdots,K\}$. The source node need to execute some computation-intensive tasks. Due to the limited computing capability, the source node needs to offload part of its computation tasks to idle helper nodes. The IRS equipped with $N$ passive reflecting elements is properly deployed to improve the task offloading efficiency. The channel coefficients from the source node to the IRS, from the source node to the $k$-th helper node, and from the IRS to the $k$-th helper node are denoted by $\mathbf{h}_r\in\mathbb{C}^{N\times 1}$, $h_{d,k}\in\mathbb{C}^{1\times 1}$and $\mathbf{g}_k^H\in\mathbb{C}^{1\times N}$, respectively.

The $D$-bit task at the source node will be partitioned into $K+1$ parts, with the $k$-th $\{k\in\mathcal{K}\}$ part of $D_k$ bits being offloaded to the $k$-th helper node and the remaining $D_0$ bits being processed by the source node locally, i.e.,
\begin{equation}
\sum\limits_{k=0}^K D_k=D.
\end{equation}
\begin{figure}[t]
\begin{center}
\includegraphics[width=0.5\textwidth]{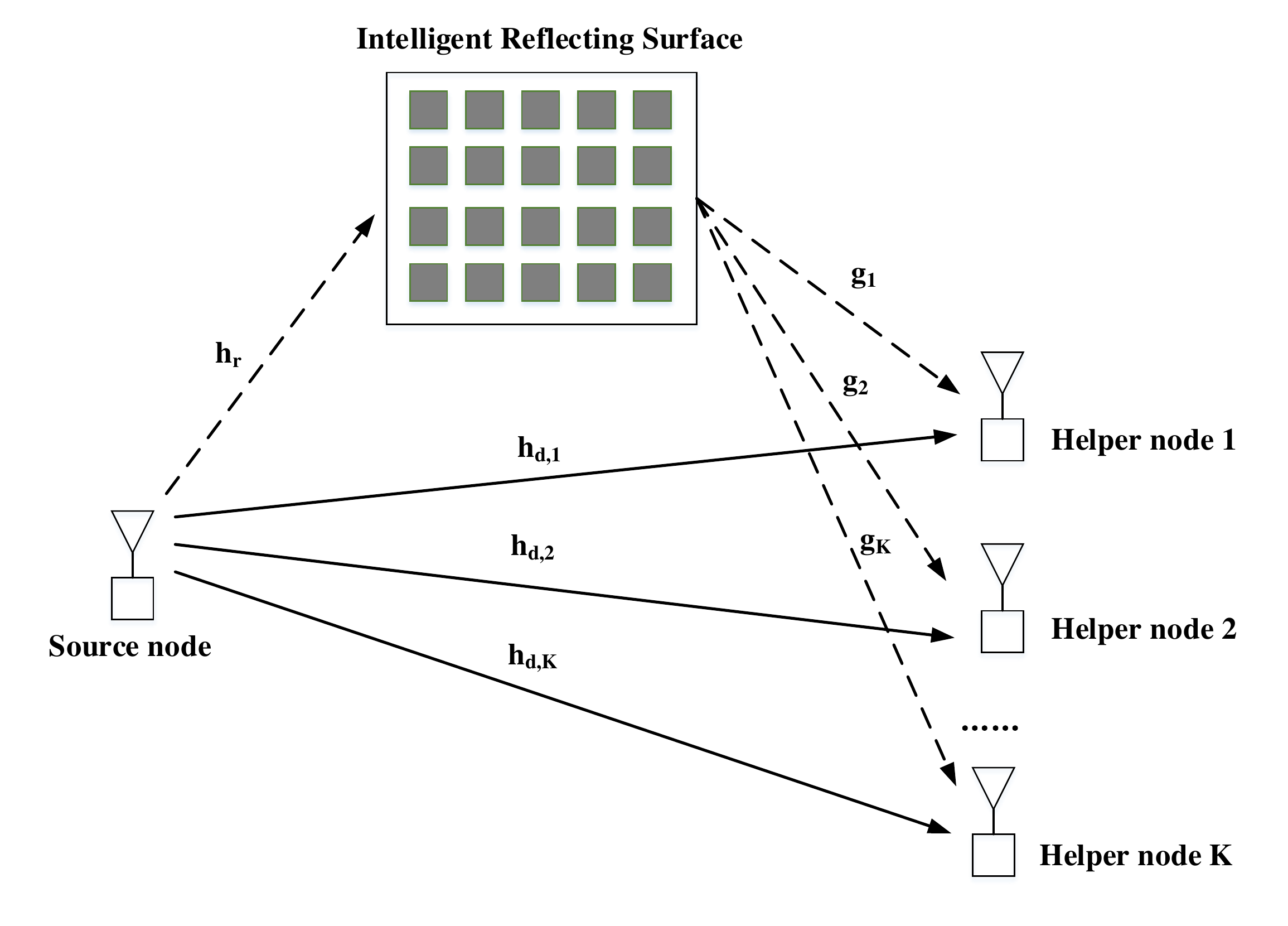}
\caption{The illustration of IRS-enhanced D2D cooperative computing.}
\end{center}
\end{figure}

Letting $f_{l,0}$ denote the CPU-cycle frequency at the source node, the computing delay for executing $D_0$ task bits locally will be
\begin{equation}
T_0 = \frac{CD_0}{f_{l,0}},
\end{equation}
where $C$ denotes the number of CPU cycles required for processing 1-bit task.

The delay of D2D cooperative computing includes two parts, namely the transmission delay and the computing delay. Since the size of the computing result is generally much smaller than that of computation task, the delay for transmitting the computing result back to the source node can be ignored \cite{7762913}, \cite{8264794}. To avoid high-complex success interference cancellation in non-orthogonal multiple access scheme, we adopt the frequency-division-multiple-access (FDMA) to support the simultaneous task offloading from the source node to multiple helper nodes.

The transmission rate from the source node to $k$-th helper node is given by
\begin{equation}
R_k=b_kB\log_2\left(1+\frac{p_k|\mathbf{g}_k^H\Lambda\mathbf{h}_r+h_{d,k}|^2}{b_kBN_0}\right),
\end{equation}
where $b_k\in[0,1]$ represents the bandwidth allocation coefficient for the $k$-th helper node, $B$ is the total bandwidth, $p_k$ is the source node transmission power toward the $k$-th helper node, $\Lambda=\text{diag}(\beta_1e^{j\theta_1},\beta_2e^{j\theta_2},\cdots,\beta_Ne^{j\theta_N})$ is the reflection coefficient matrix, where $\beta_n\in[0, 1]$ and $\theta_n\in[0,2\pi]$ are the reflection amplitude and phase shift of the $n$-th passive element, respectively, and $N_0$ denotes the additive Gaussian noise power spectral density. We set $\beta_n=1$ to maximize the reflected signal power.

Therefore, the delay for D2D cooperative computing at the $k$-th helper node will be
\begin{equation}
T_k=\frac{D_k}{R_k}+\frac{D_kC}{f_{l,k}},
\end{equation}
where $f_{l,k}$ is the CPU-cycle frequency of the $k$-th helper node.


In this paper, we aim at minimizing the task execution delay by jointly optimizing the computation task assignment $\{D_k\}$, bandwidth allocation $\{b_k\}$, transmission power $\{p_k\}$ and IRS phase beamfroming $\Lambda$, the optimization problem can be formulated as
\begin{subequations}
\begin{align}
\underset{\{D_k,b_k,p_k\},\Lambda}{\text{minimize }}&~~\underset{k\in\{0,\mathcal{K}\}}{\text{max}} \{T_k\}\\
\text{s.t. }~~~~~&\sum\limits_{k=0}^K D_k=D,\\
& \sum\limits_{k=1}^K b_k\leq 1,\\
&\sum\limits_{k=1}^K p_k\leq P_{\text{max}},\\
&0\leq\theta_n\leq 2\pi, \forall n\in\{1,2,\cdots,N\},\\
&D_k,b_k,p_k\geq 0,
\end{align}
\end{subequations}
where (5b) is the task allocation constraint, (5c) denotes the bandwidth allocation constraint, (5d) restricts the maximum transmission power $P_{\text{max}}$ at the source node, and (5e) gives the constraints on phase beamforming of IRS.

It is worth noting that problem (5) is non-convex and is difficult to solve, due to the following reasons: (i) the min-max formulation; (ii) the fractional expression of delay in (4); and (iii) the transmit power and phase beamforming variables coupled in the objective function.
\section{Proposed Method}
In order to address the non-convex problem (5), we first introduce an auxiliary variable $t=\underset{k\in\{0,\mathcal{K}\}}{\text{max}} \{T_k\}$ to transform the min-max problem as follows
\begin{subequations}
\begin{align}
\underset{\{D_k,b_k,p_k,f_{l,k}\},\Lambda, t}{\text{minimize }}&~~t\\
\text{s.t. }~~~~~&T_k\leq t,\forall k\in\{0,\mathcal{K}\},\\
&\text{(5b)-(5f)}.
\end{align}
\end{subequations}

Next, we adopt the block coordinate descent (BCD) method to transform the non-convex problem (6) into three subproblems, namely the computation task assignment, transmit power and bandwidth allocation, and phase beamforming optimization subproblems.
\subsection{Computation Task Assignment}
For given $\{b_k^\ast,p_k^\ast,\Lambda^\ast\}$, (6) reduces to the following computation task assignment subproblem:
\begin{subequations}
\begin{align}
\underset{\{D_k\}, t}{\text{minimize }}&~~t\\
\text{s.t. }~~~~~&T_0=\frac{CD_0}{f_{l,0}}\leq t,\\
\begin{split}&T_k=\frac{D_k}{b_k^\ast B\log_2\left(1+\frac{p_k^\ast|\mathbf{g}_k^H\Lambda^\ast\mathbf{h}_r+h_{d,k}|^2}{b_k^\ast BN_0}\right)}\\
&~~~~~~+\frac{D_kC}{f_{l,k}}\leq t, \forall k\in\mathcal{K},\end{split}\\
&\text{(5b), (5f)}.
\end{align}
\end{subequations}
We can see that (7) is a linear programming (LP) problem. By exploiting the convex optimization theory, we further derive the optimal solution of task assignment in \emph{Proposition 1}.

\emph{Proposition 1:} The optimal task assignment strategy of (7) satisfies
\begin{equation}
D_k^\ast=\frac{x_kD}{\sum\limits_{k=0}^Kx_k},k\in\{0,\mathcal{K}\},
\end{equation}
where $x_0=\frac{f_{l,0}}{C}$ and $x_k=\frac{1}{\frac{1}{R_k^\ast}+\frac{C}{f_{l,k}}},k\in\mathcal{K}$.

\emph{Proof:} See Appendix A.
\subsection{Transmit Power and Bandwidth Allocation}
Under given $\{D_k^\ast, \Lambda^\ast\}$, (6) reduces to the transmit power and bandwidth allocation subproblem, which can be expressed as
\begin{subequations}
\begin{align}
\underset{\{b_k,p_k\}, t}{\text{minimize }}&~~t\\
\begin{split}\text{s.t. }&b_k B\log_2\left(1+\frac{p_k|\mathbf{g}_k^H\Lambda^\ast\mathbf{h}_r+h_{d,k}|^2}{b_k BN_0}\right)\\
&\geq \frac{D_k^\ast}{t-\frac{D_k^\ast C}{f_{l,k}}},\forall k\in\mathcal{K},\end{split}\\
&\text{(5c)-(5d), (5f), (7b)},
\end{align}
\end{subequations}
where (9b) is derived from the constraint (6b), i.e., $T_k\leq t$. Based on the convex optimization theory, $B\log_2\left(1+\frac{p_k|\mathbf{g}_k^H\Lambda^\ast\mathbf{h}_r+h_{d,k}|^2}{BN_0}\right)$ is a concave function with respect to $p_k$, thus its perspective function $b_k B\log_2\left(1+\frac{p_k|\mathbf{g}_k^H\Lambda^\ast\mathbf{h}_r+h_{d,k}|^2}{b_k BN_0}\right)$ is also concave. Moreover, $\frac{D_k^\ast}{t-\frac{D_k^\ast C}{f_{l,k}^\ast}}$ is a convex function of $t$. Therefore, the constraint (9b) is convex. Since the other constraints and the objective function is linear, problem (9) is a convex optimization problem, and classic convex optimization algorithms, e.g., the interior-point method, can be applied to solve it.
\subsection{Phase Beamforming Optimization}
For Given $\{D_k^\ast, p_k^\ast, b_k^\ast\}$, (6) reduces to the following phase beamforming optimization subproblem:
\begin{subequations}
\begin{align}
\underset{\Lambda, t}{\text{minimize }}&~~t\\
\begin{split}\text{s.t. }&b_k^\ast B\log_2\left(1+\frac{p_k^\ast|\mathbf{g}_k^H\Lambda\mathbf{h}_r+h_{d,k}|^2}{b_k^\ast BN_0}\right)\geq \\
&\frac{D_k^\ast}{t-\frac{D_k^\ast C}{f_{l,k}}},\forall k\in\mathcal{K},\end{split}\\
&\text{(5e), (7b)}.
\end{align}
\end{subequations}

We set $\mathbf{v}=(\beta_1e^{j\theta_1},\beta_2e^{j\theta_2},\cdots,\beta_Ne^{j\theta_N})^T$, $\hat{\mathbf{v}}=[\mathbf{v}^T, 1]^T$, and $[\mathbf{g}_k^H\text{diag}(\mathbf{h}_r),h_{d,k}]=\mathbf{h}_k^H$. Since $|\mathbf{g}_k^H\Lambda\mathbf{h}_r+h_{d,k}|=[\mathbf{g}_k^H\text{diag}(\mathbf{h}_r),h_{d,k}]\hat{\mathbf{v}}=\mathbf{h}_k^H\hat{\mathbf{v}}$, (10) can be transformed to the following problem:
\begin{subequations}
\begin{align}
\underset{\hat{\mathbf{v}}, t}{\text{minimize }}&~~t\\
\text{s.t. }&b_k^\ast B\log_2\left(1+\frac{p_k^\ast|\mathbf{h}_k^H\hat{\mathbf{v}}|^2}{b_k^\ast BN_0}\right)\geq \frac{D_k^\ast}{t-\frac{D_k^\ast C}{f_{l,k}}},\forall k\in\mathcal{K},\\
&[\hat{\mathbf{v}}\hat{\mathbf{v}}^H]_{nn}=1, \\
&\text{(7b)}.
\end{align}
\end{subequations}
In order to tackle the second-order equality constraint (11c), we introduce a matrix variable $\hat{\mathbf{V}}=\hat{\mathbf{v}}\hat{\mathbf{v}}^H$, thus (11) is equivalent to
\begin{subequations}
\begin{align}
\underset{\hat{\mathbf{V}},t}{\text{minimize }}&~~t\\
\text{s.t. }&b_k^\ast B\log_2\left(1+\frac{p_k^\ast\text{Tr}(\mathbf{H}_k\hat{\mathbf{V}})}{b_k^\ast BN_0}\right)\geq \frac{D_k^\ast}{t-\frac{D_k^\ast C}{f_{l,k}}},\forall k\in\mathcal{K},\\
&[\hat{\mathbf{V}}]_{nn}=1, \forall k\in\mathcal{K},\forall n\in\{1,2,\cdots,N+1\},\\
&\hat{\mathbf{V}}\succeq 0, \\
&\text{Rank}(\hat{\mathbf{V}})=1,\\
&\text{(7b)},
\end{align}
\end{subequations}
where $\mathbf{H}_k=\mathbf{h}_k\mathbf{h}_k^H$. Then, we adopt the SDR method to relax the rank-one constraint (12e). As a result, the optimal solution of problem (12) cannot be ensured rank-one, which implies that the optimal value of problem (12) will be an upper bound of problem (10). 

According to above derivations, we develop an alternating optimization method for solving (5). All the variables in (5) will be divided into three blocks, namely ($\{D_k\},\{b_k,p_k\},\hat{\mathbf{V}}$). Then, the task assignment $\{D_k\}$, the transmit power and bandwidth allocation $\{b_k,p_k\}$, and the IRS phase beamforming design $\hat{\mathbf{V}}$ are alternately optimized by solving (7), (9), and (12), respectively, while ensuring the other variables fixed. The detailed procedure is illustrated in Algorithm 1.
\begin{algorithm}[htbp]
\caption{Proposed Method for Solving (5)}
\textbf{Initialize:} Set ($D_k^{(i)},b_k^{(i)},p_k^{(i)},\Lambda^{(i)}$), and $i=1$.\\
\textbf{Repeat:}\\
\quad Calculate the optimal computation task assignment scheme $D_k^{(i)}$ according to \emph{Proposition 1};\\
\quad Solve (9) to obtain optimal transmit power and bandwidth allocation strategy ($b_k^{(i)},p_k^{(i)}$);\\
\quad Obtain the optimal $\hat{\mathbf{V}}^{(i)}$ by solving (12);\\
\quad Update the iterative number $i=i+1$;\\
\textbf{Until} the increase of objective function is below a threshold $\epsilon$.\\
\textbf{Recover} the optimal rank-one phase beamforming $\Lambda^\ast$ from the optimal $\hat{\mathbf{V}}^\ast$ by utilizing the Gaussian randomization method.\\
\textbf{Return } optimal solution ($D_k^\ast,b_k^\ast,p_k^\ast,\Lambda^\ast$).\\
\end{algorithm}
\subsubsection{Convergence Analysis} To analyze the convergence of Algorithm 1, we first prove that the objective function $t(D_k,b_k,p_k,\mathbf{V})$ is non-increasing after each iteration.
It follows that
\begin{equation}
\begin{aligned}
&t(D_k^{(i)},b_k^{(i)},p_k^{(i)},\mathbf{V}^{(i)})\overset{(a_1)}{\geq} t(D_k^{(i+1)},b_k^{(i)},p_k^{(i)},\mathbf{V}^{(i)})\overset{(a_2)}{\geq}\\
&t(D_k^{(i+1)},b_k^{(i+1)},p_k^{(i+1)},\mathbf{V}^{(i)})\overset{(a_3)}{\geq}\\
&t(D_k^{(i+1)},b_k^{(i+1)},p_k^{(i+1)},\mathbf{V}^{(i+1)}),
\end{aligned}
\end{equation}
where $(a_1)$, $(a_2)$ and $(a_3)$ are due to the optimal value of $t$ obtained from (7), (9), (12), respectively. Besides, the total computing delay is lower bounded by a certain value, thus the proposed Algorithm 1 can converge to the optimal solution.
\section{Performance Evaluation}
In this section, extensive numerical results are provided to validate the performance of our proposed IRS enhanced D2D cooperative computing strategy. For comparison, we also present the following three benchmark methods:
\begin{itemize}
\item \emph{Partial offloading without IRS}: the source node offloads part of its computation task to helper nodes without the aid of IRS.
\item \emph{Full offloading only}: the source node offloads all of its computation task to helper nodes, i.e., $D_0=0$.
\item \emph{Local computing only}: all computation tasks at the source node will be executed by local computing, thus the computing delay is calculated as $T=\frac{CD}{f_{l,0}}$.
\end{itemize}

\begin{figure*}[!htb]
\minipage{0.33\textwidth}%
  \includegraphics[width=\linewidth]{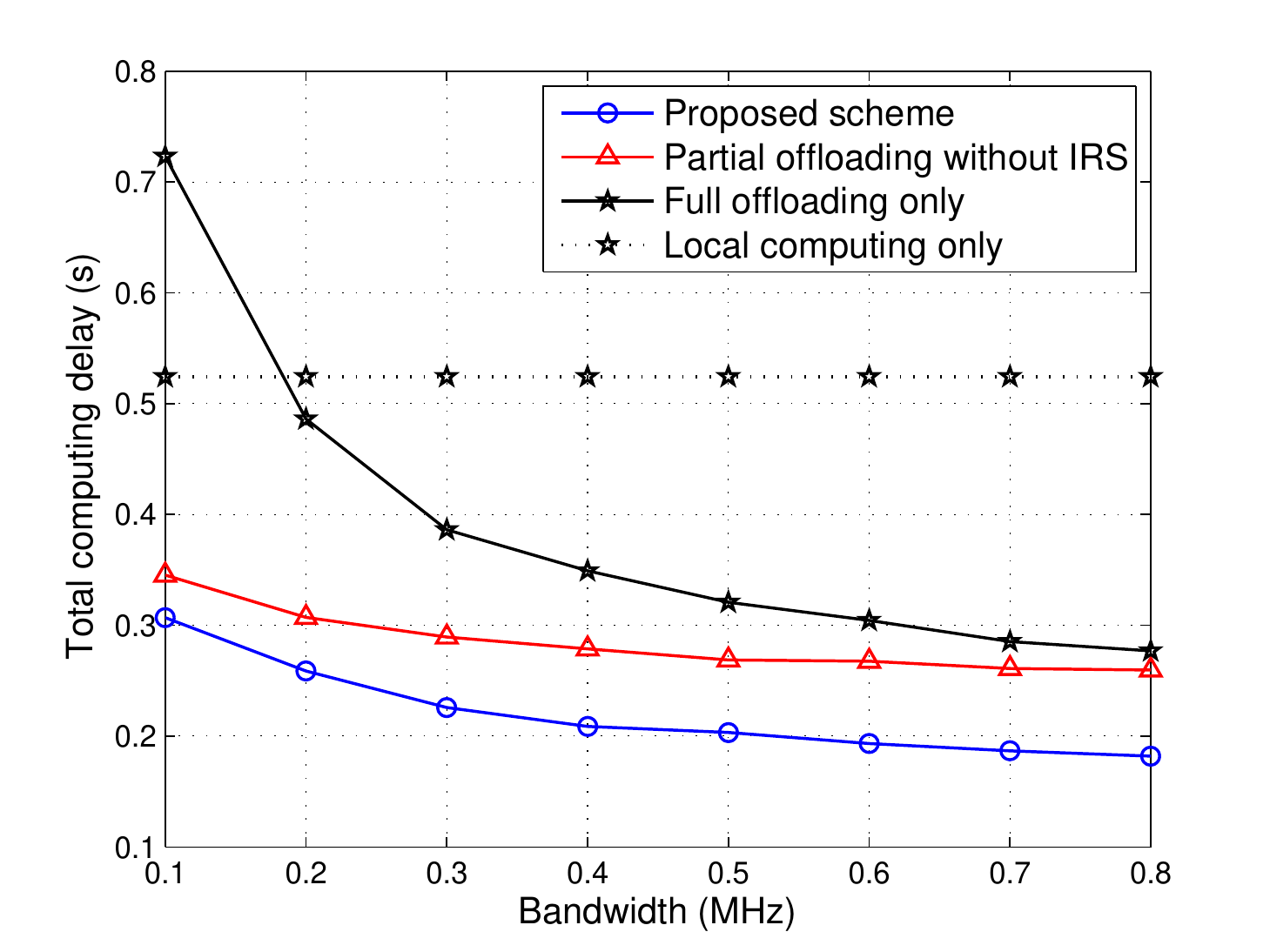}
  \caption{Total computing delay versus \\bandwidth.}
\endminipage
\minipage{0.33\textwidth}
  \includegraphics[width=\linewidth]{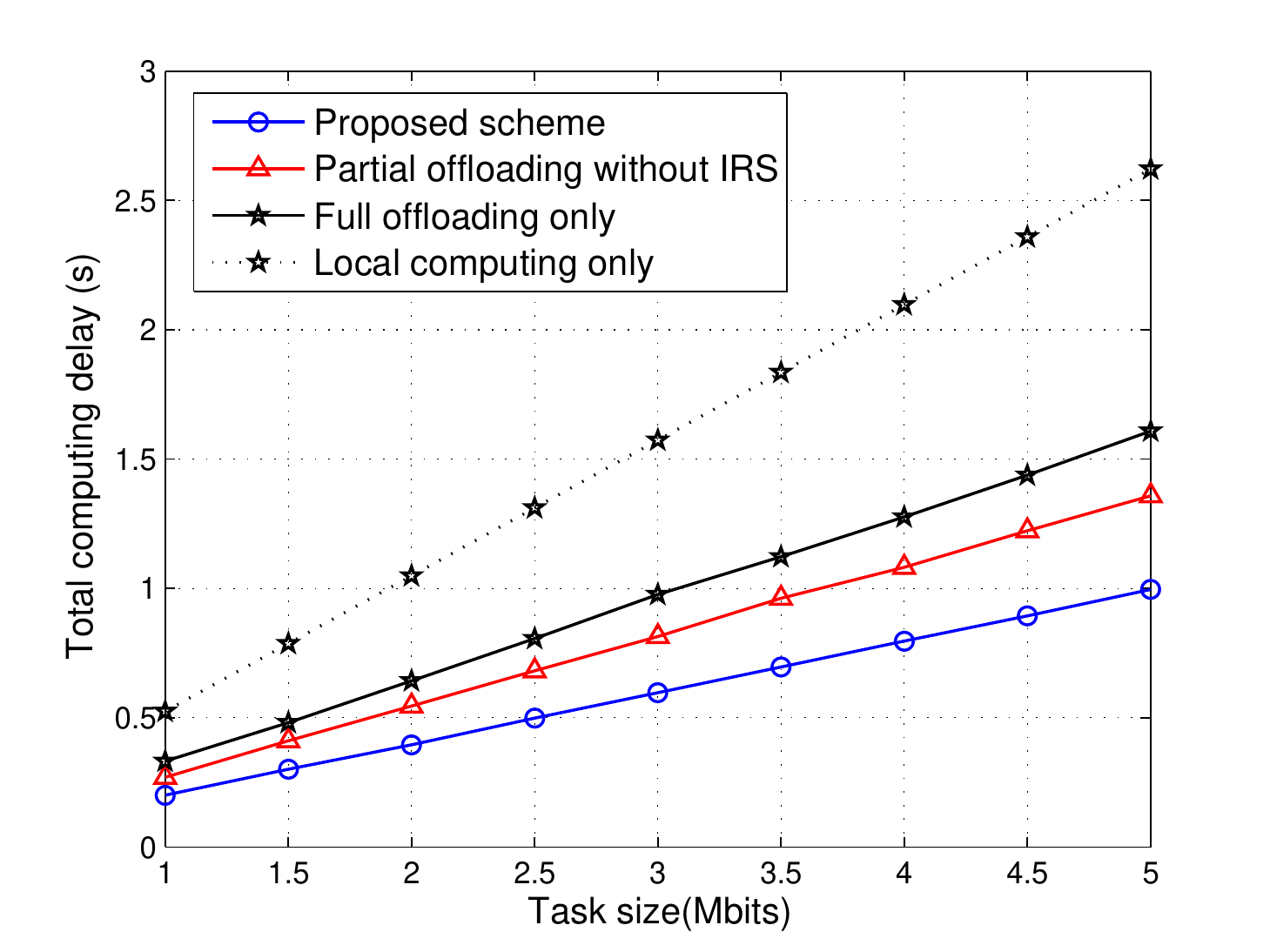}
  \caption{Total computing delay versus \\task size.}
\endminipage\hfill
\minipage{0.33\textwidth}%
  \includegraphics[width=\linewidth]{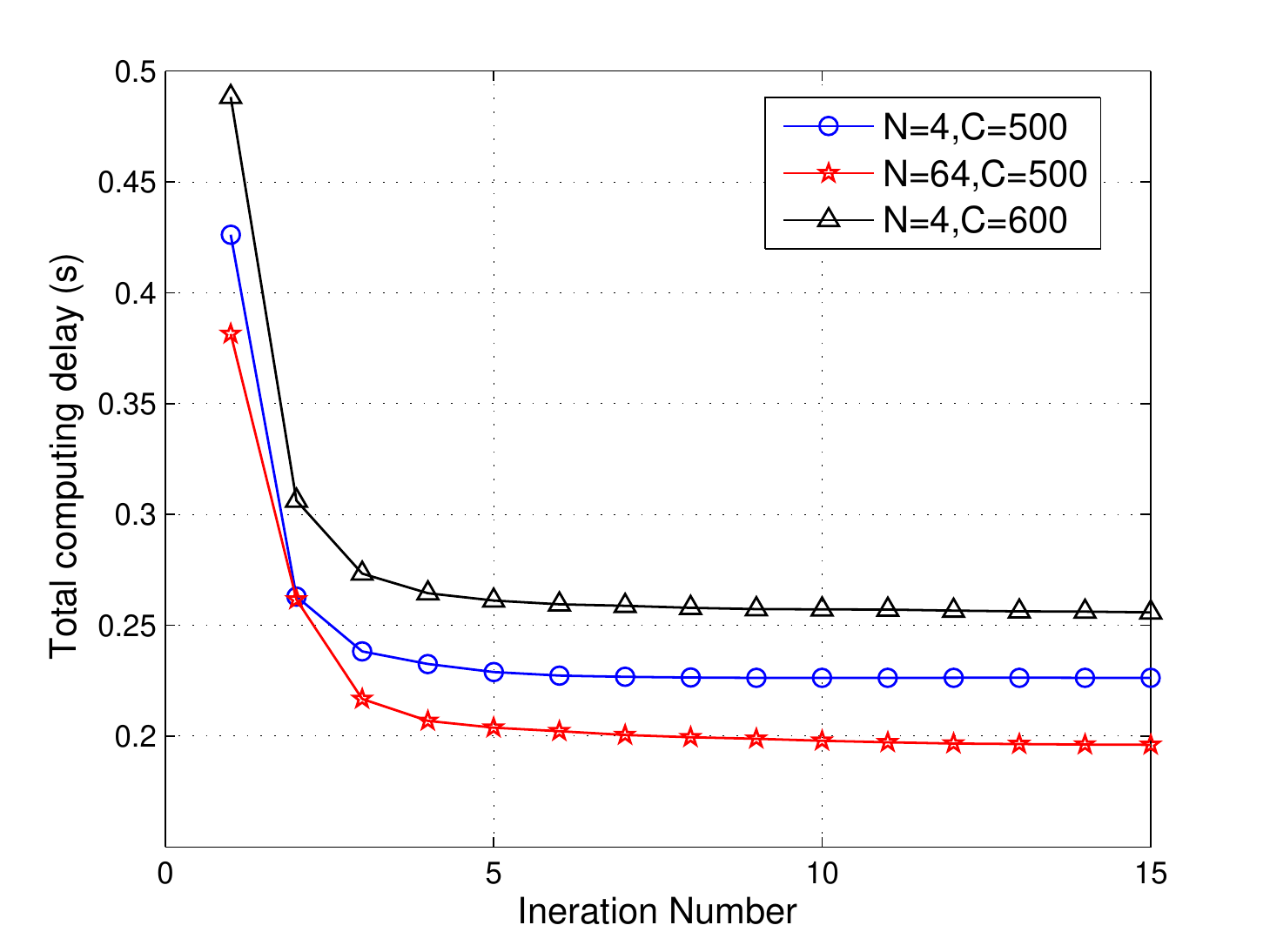}
  \caption{Total computing delay versus \\iteration number.}
\endminipage
\end{figure*}

In the simulations, the number of helper nodes is set as $K=2$, the source node is located at the original point, IRS is located at (0,5), and the helper nodes are located at the horizontal coordinates [(1,5), (2,4)]. The large-scale fading is given by $L(d)=C_0d^{-\alpha}$, where $C_0=-30$ dB and $\alpha=3$. The small-scale fading follows the Rayleigh fading with an unity mean. In addition, it is assumed that the direct link between the source node and $2$-th helper node is fully blocked. The other simulation parameters are set as follows: $N_0=10^{-16}$ Watt/Hz, $P_\text{max}=1$ Watt, $f_{l,0}=1$ GHz, $f_{l,1}=1.2$ GHz, and $f_{l,2}=1.5$ GHz \cite{7762913}, \cite{8264794}.

In Fig. 2, we plot the total computing delay versus system bandwidth with $N=32$ and $D=1$ Mbits,. It is observed that the total computing delay decreases with the increase of system bandwidth. This is due to the fact that a larger bandwidth implies a higher offloading rate.
As compared to the partial offloading scheme without the aid of IRS, our proposed method can achieve lower computing delay. It demonstrates that IRS plays a vital role in improving the user experience in terms of computing delay. Another important observation is that our proposed scheme exhibits a larger delay reduction than the full offloading strategy, when the bandwidth is small. This is because that the limited bandwidth restricts the task offloading rate, and further incurs a high total computing delay of full offloading scheme. It also reveals that the local computing is more efficient than computing offloading in the communication-restricted scenario. Meanwhile, we see that the proposed scheme can achieve lower computing delay than the local computing scheme, and the gap between them larger as the bandwidth increases.

Fig. 3 shows the total computing delay versus the size of computation tasks with $N=32$ and $B=0.5$ MHz. As desired, the total computing delay increases monotonically with the size of the computation tasks. Obviously, a larger size of the computation tasks will cause higher task execution delay and longer task offloading duration. Furthermore, we also observe that the proposed scheme outperforms the baseline methods in delay reduction, especially when the size of task is large. In Fig. 4, we reveal the convergence rate of the proposed Algorithm 1 with $B=0.5$ MHz and $D=1$ Mbits. As can be observed, the proposed method will converge to the optimal point within a few iterations. Moreover, it is observed that the computing delay decreases with the number of reflection elements at IRS. This is because that a larger number of reflections elements implies a higher freedom to design phase beamforming for improving the task transmission rate. Finally, we see that a higher task computational complexity $C$ will result in a larger computing delay.
\section{Conclusion}
This paper studied the IRS-assisted D2D cooperative computing. We aimed at minimizing the total computing delay with the joint optimization of computation task assignment, transmit power, bandwidth allocation, and phase beamforming design. Furthermore, we developed an alternating optimization method to solve the formulated problem, while the optimal computation task assignment was derived in closed-form expression and the phase beamforming was obtained by utilizing SDR method. Finally, our simulation results validated the performance gain of the proposed IRS-assisted D2D cooperative computing strategy compared to other benchmark methods.

\section*{Appendix A: Proof of \emph{Proposition 1}}
The Karush-Kuhn-Tucker (KKT) conditions can be applied to obtain the optimal solution of (7). The Lagrange function of (7) is given by
\begin{equation}
\begin{aligned}
&L(D_k,t,\lambda_k,\mu)=t+\lambda_0\left(\frac{CD_0}{f_{l,0}}-t\right)+\\
&\sum\limits_{k=1}^K\lambda_k\left(\frac{D_k}{R_k^\ast}+\frac{D_kC}{f_{l,k}}-t\right)\mu\left(\sum\limits_{k=0}^KD_k-D\right),
\end{aligned}
\end{equation}
where $R_k^\ast=b_k^\ast B\log_2\left(1+\frac{p_k^\ast|\mathbf{g}_k^H\Lambda^\ast\mathbf{h}_r+h_{d,k}|^2}{b_k^\ast BN_0}\right)$, $\lambda_k$ and $\mu$ denote the dual variables associated with the inequality and equality constraints of (7), respectively. The KKT conditions will be
\begin{equation}
\frac{\partial L}{\partial t}=1-\sum\limits_{k=0}^K\lambda_k^\ast=0,~\frac{\partial L}{\partial D_0}=\frac{C\lambda^\ast_0}{f_{l,0}}+\mu^\ast=0,
\end{equation}
\begin{equation}
\frac{\partial L}{\partial D_k}=(\frac{1}{R_k^\ast}+\frac{C}{f_{l,k}})\lambda_k^\ast+\mu^\ast=0,k\in\mathcal{K},
\end{equation}
\begin{equation}
\lambda_0^\ast\left(\frac{CD_0^\ast}{f_{l,0}}-t^\ast\right)=0,~\lambda_k^\ast\left(\frac{D_k^\ast}{R_k^\ast}+\frac{D_k^\ast C}{f_{l,k}}-t^\ast\right)=0,k\in\mathcal{K},
\end{equation}
\begin{equation}
\sum\limits_{k=0}^K D_k^\ast=D,~\lambda_k^\ast\geq 0,k\in\{0,\mathcal{K}\},
\end{equation}
where (15)-(16) denote the first-order derivative optimality conditions, (17) indicates the complementary slackness conditions, and (18) is the primal/dual feasible conditions.

According to (15)-(16) and (18), we have $\lambda_k^\ast>0$. Then, we can derive the following equality based on (17):
\begin{equation}
t^\ast=\frac{D_0^\ast}{x_0}=\frac{D_1^\ast}{x_1}=\cdots=\frac{D_K^\ast}{x_K}.
\end{equation}
Applying the equality $\sum\limits_{k=0}^K D_k^\ast=D$, we complete the proof.
\bibliography{main}
\end{document}